\documentclass[twocolumn,secnumarabic,amssymb, showpacs, nobibnotes, aps, prd]{revtex4-1}

\usepackage{graphicx}
\usepackage{subfigure}
\usepackage{ae} 
\usepackage{float}  
\usepackage{amsmath}
\usepackage{color}

\begin{document}

\preprint{APS/123-QED}

\title{Observation of the nonlinear Wood's anomaly\\ on periodic arrays of nickel nanodimers}

\author{Ngoc-Minh Tran$^{1}$}
\author{Ioan-Augustin Chioar$^{2}$}
\author{Aaron Stein$^{3}$}
\author{Alexandr Alekhin$^{1}$}
\author{Vincent Juv\'e$^{1}$}
\author{Gwena\"elle Vaudel$^{1}$}
\author{Ilya Razdolski$^{1,4}$}
\author{Vassilios Kapaklis$^{2}$}
\author{Vasily Temnov$^{1,}$}
\email[]{vasily.temnov@univ-lemans.fr}

\affiliation{$^1$Institut des Mol\'ecules et Mat\'eriaux du Mans, UMR CNRS 6283, Le Mans Universit\'e, 72085 Le Mans, France}%
\affiliation{$^2$Department of Physics and Astronomy, Uppsala University, 751 20 Uppsala, Sweden}%
\affiliation{$^3$Center for Functional Nanomaterials, Brookhaven National Laboratory, Upton, New York 11973, USA}%
\affiliation{$^4$Fritz-Haber-Institut der MPG, Phys. Chemie, Faradayweg 4-6, 14195 Berlin, Germany}%


\date{\today}

\begin{abstract}
Linear and nonlinear magneto-photonic properties of periodic arrays of nickel nanodimers are governed by the interplay of the (local) optical response of individual nanoparticles and (non-local) diffraction phenomena, with a striking example of Wood's anomaly. Angular and magnetic-field dependencies of the second harmonic intensity evidence Wood's anomaly when new diffraction orders emerge. Near-infrared spectroscopic measurements performed at different optical wavelengths and grating constants discriminate between the linear and nonlinear excitation mechanisms of Wood's anomalies. In the nonlinear regime the Wood's anomaly is characterized by an order-of-magnitude larger effect in intensity redistribution between the diffracted beams, as compared to the linear case. The nonlinear Wood's anomaly manifests itself also in the nonlinear magnetic contrast highlighting the prospects of nonlinear magneto-photonics. 
\end{abstract}

\pacs{Valid PACS appear here}
\maketitle


\section{\label{sec1:intro}Introduction}

Combining functionalities of plasmonic materials~\cite{Ebbesen1998} with optical properties of diffraction gratings and periodic structures with a sub-wavelength periodicity, i.e. optical meta-surfaces~\cite{Auguie2008,Kravets2008,zhounatnano2011,czaplicki2016,Michaeli2017,Kuzmin2018}, represents a topical area in nanophotonics. 
These systems are often discussed in the context of  plasmonics~\cite{barnes2003,lezec2004,genet2010} and magneto-plasmonics ~\cite{wurtz2008,Torrado2010,belotelov2011,Chen2013,chekhov2018acs}, if one of the ingredients displays magneto-optical properties. Nonlinear-optical studies highlight the role of propagating surface plasmon polaritons (SPPs) excited on silver gratings~\cite{Quail:88}, thin noble metal films \cite{Novotny2008,Grosse2012} or noble metal-ferromagnet multilayer structures \cite{razdolski2016}.

In plasmonics of more sophisticated periodic arrangements of metallic nanostructures, spectrally narrow resonances, commonly referred to as surface lattice resonances (SLRs)~\cite{Kravets2008,zhounatnano2011,zhou2012acs,kataja2015}, arise from an unusual interplay between the localized surface plasmon resonance (LSPR) and the emerging diffraction orders under the condition of Wood's anomaly. The latter describes the redistribution of light intensity between different diffracted beams upon opening of a new diffraction order propagating along the grating surface \cite{wood1902}.

In magneto-plasmonics, the excitation of Wood's anomalies in two-dimensional arrays of ferromagnetic nanodisks results in the enhancement of the magneto-optical Kerr effect (MOKE)~\cite{chetvertukhin2012,kataja2015}. In these structures, plasmonic effects are enabled either by a high relative content of metal, resulting in the excitation of SPPs despite the high optical losses in ferromagnetic transition metals, or LSPRs in nanosized metallic objects.

A conventional diffraction grating requires the grating period $\Lambda$ to be larger than a half of an optical wavelength $\lambda$ ($\Lambda>\lambda/2$) to ensure the presence of at least one diffraction order. In contrast, optical meta-surfaces are usually characterized by a much smaller periodicity $\Lambda\ll\lambda$~\cite{Yu2014, Kildishev2013, Kuzmin2018}. Overcoming this frontier, an intriguing crossover regime can be explored within the realms of nonlinear optics, where the wavelength $\lambda$ of the fundamental radiation is converted to $\lambda/n$ by means of a nonlinear-optical process of the order $n>1$. In this case, the same periodic structure can serve as a meta-surface for the fundamental wavelength $\lambda$, and as a regular grating for frequency-converted radiation at a shorter wavelength $\lambda/n$. To the best of our knowledge, the nonlinear-optical properties in this intriguing transition regime remain up to date unexplored.

In this work, we employ the nonlinear-optical technique of magneto-induced second harmonic (SH) generation to study the Wood's anomaly, i.e. the intrinsic property of an optical grating in the aforementioned nonlinear transition regime. We perform angle-dependent SH spectroscopy with a tunable femtosecond laser source on a rectangular array of nickel dimers featuring a nanoscale gap, similar to that studied in Ref.~\cite{Kravets2008}.
Taking advantage of different periods of the investigated array in two orthogonal in-plane directions, we identify a novel regime for the nonlinear Wood's anomaly, i.e. when the structure exhibits grating properties exclusively at the SH wavelength $\lambda/2$ but not at the fundamental wavelength $\lambda$.  Our experiments represent the first step to understand the transition between optical diffraction gratings and meta-surfaces, extending the concept of Wood's anomaly~\cite{wood1935,maystre2012book} to the nonlinear regime. Our results are inline with theoretical predictions~\cite{DadoenkovaAPL1999} that nonlinear-optical and magneto-optical effects can be utilized to characterize the efficiency of diffraction gratings in periodic arrays of magnetic nanostructures.

\begin{figure*}
\centering
\includegraphics[width=1\textwidth]{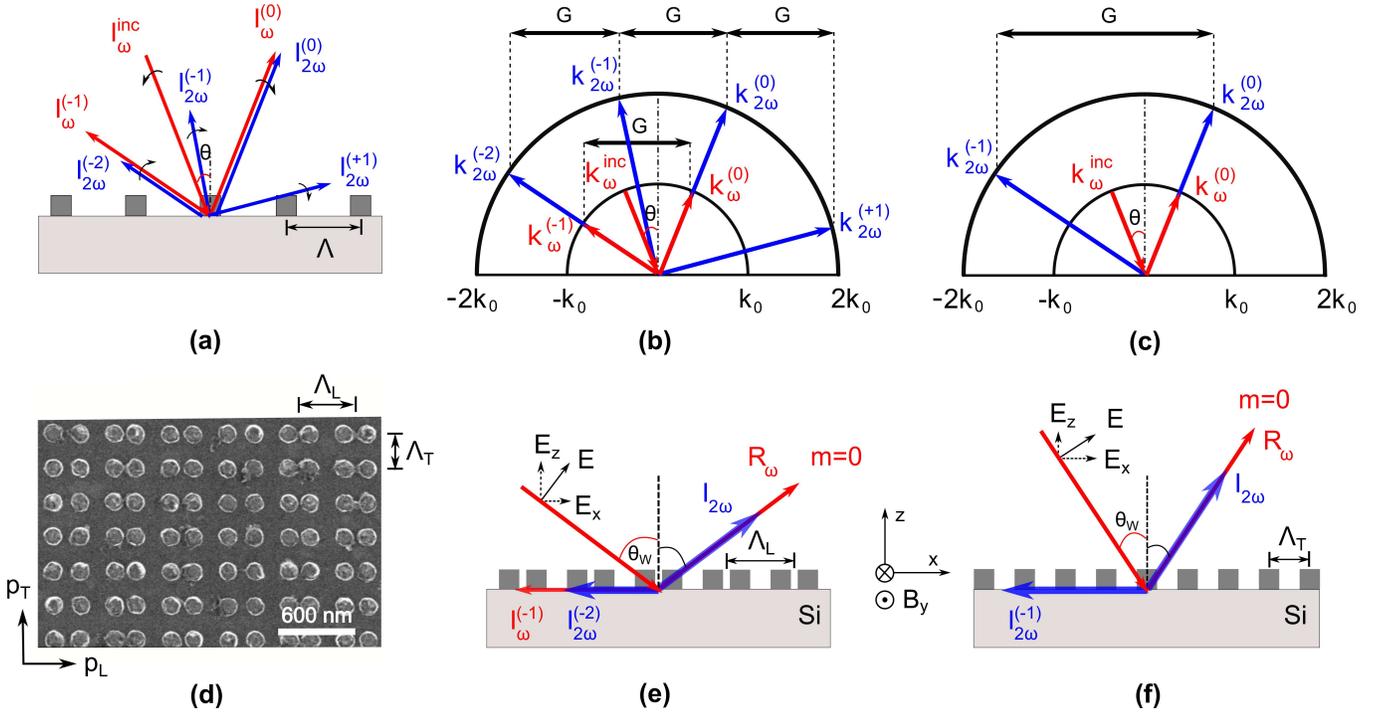}
\caption{
(a) Linear (red) and  nonlinear SH (blue) diffraction orders in reflection from a grating with a periodicity $\Lambda>\lambda/2$. When the angle of incidence of the fundamental radiation $I^{inc}_\omega$ increases in the counter-clockwise direction, all diffraction orders rotate clock-wise. 
(b-c) In the reciprocal space the in-plane wavevector components of linear and nonlinear diffraction orders are shifted by a multiple of the grating wavevector $G=2\pi/\Lambda$. 
(b) Grating regime. For $G\leq2k_0$ or $\Lambda \geq \lambda/2$ both linear and nonlinear SH diffraction are allowed. 
(c) Transition regime. When $2k_0< G<4k_0$ or $\lambda/4<\Lambda<\lambda/2$ linear diffraction is suppressed and only the SH nonlinear diffraction order $m_{nl}=-1$ is allowed.
(d) Scanning electron microscopy image of the rectangular array of nickel nanodimers on silicon. 
(e-f) Excitation geometries for Wood's anomalies. (e) Co-excitation of linear and nonlinear Wood's anomalies: for large periodicity $\Lambda_L=$~445~nm, both linear $I_\omega^{(-1)}$ and nonlinear $I_{2\omega}^{(-2)}$ in-plane diffraction orders are excited at $\theta=\theta_W$. (f) Purely nonlinear Wood's anomaly: for small periodicity $\Lambda_T=$~265~nm, only the non-linear diffraction order $I_{2\omega}^{(-1)}$ emerges at $\theta=\theta_W$.}
\label{fig:SHG_diffraction}
\end{figure*}

\section{Linear and nonlinear diffraction on optical gratings}

Diffraction on a grating in the linear and nonlinear-optical regimes
is illustrated in Fig.~\ref{fig:SHG_diffraction}. Let us consider the case of a regular grating with a periodicity $\Lambda>\lambda/2$ irradiated at an angle of incidence $\theta$ (Fig.~\ref{fig:SHG_diffraction}a). The propagation direction of a diffracted beam of an order $m$ is determined by the phase-matching condition, which is conventionally written for in-plane ($x$-) projections of the wavevectors:
\begin{equation} \label{linear_PM}
k_x^{out}(\omega)=k_x^{in}(\omega)+mG\,,
\end{equation}
where $k_x^{in}(\omega)=k_0\sin\theta$, $k_0=2\pi/\lambda$ and $G=2\pi/\Lambda$ is the grating constant in the reciprocal space. Imagine that the intensity of incident light is so high that the grating itself converts the frequency of incident light in the nonlinear optical process, for example it doubles the frequency by SH generation at surfaces and interfaces. 
In the electric dipole approximation, the generic second-order nonlinear polarization $\mathbf{P}(2\omega)$ acts as a source of radiation at the double frequency $2\omega$:
\begin{equation} \label{qn:shg}
P_i(2\omega)=\varepsilon_0 \chi^{(2)}_{ijk}E_j(\omega)E_k(\omega)\,.
\end{equation}
Here $\varepsilon_0$ is the vacuum permittivity, $\chi^{(2)}_{ijk}$ is the second-order susceptibility tensor, $E_j(\omega)$ and $E_k(\omega)$ are the electric components at the fundamental frequency $\omega$; indices $i,j,k$ denote the Cartesian coordinates. At magnetic surfaces and interfaces the nonlinear susceptibility contains non-magnetic (even) and magnetization-dependent (odd) $\chi^{(2)}$-contributions~\cite{pan1989} resulting in the dependence of SH intensity on the magnetic field, see next section for details. 
Both $E_j(\omega)$ and $E_k(\omega)$ are proportional to $\exp[ik_x^{in}(\omega)x]$ resulting in $E_j(\omega)E_k(\omega)\propto\exp[2ik_x^{in}(\omega)x]$. 
Therefore, the phase-matching condition for the nonlinear SH diffraction reads~\cite{Quail:88}: 
\begin{equation} \label{nonlinear_PM}
k_{x}^{out}(2\omega)=2k_{x}^{in}(\omega)+m_{nl}G\,,
\end{equation}
where we have introduced another integer $m_{nl}\neq m$ indicating the nonlinear diffraction order. In what follows, we shall use the symbols $m_{lin}\equiv m$ and $m_{nl}$ when talking about the linear and nonlinear diffraction orders, respectively. Phase-matching conditions for linear and nonlinear diffraction can be visualized in the reciprocal space (Fig.~\ref{fig:SHG_diffraction}b). The number of nonlinear SH diffraction orders is larger than in the linear case. Linear diffraction orders are collinear with nonlinear diffraction orders. For example, the first linear diffraction order $m_{lin}=-1$ is collinear with the SH diffraction order $m_{nl}=2m_{lin}=-2$. The collinearity of linear and nonlinear diffraction orders suggests that Wood's anomaly in both regimes should occur at the same angle of incidence.

The situation becomes different when the spatial periodicity is reduced to $\Lambda<\lambda/2$  (Fig.~\ref{fig:SHG_diffraction}c). In this case, linear diffraction with $m_{lin}\neq 0$ is forbidden because of $G>2k_0$. At the same time, if $\Lambda>\lambda/4$, the phase-matching condition can be fulfilled in the nonlinear regime, resulting in the nonlinear diffraction of the order $m_{nl}=-1$. It is seen that the periodic structure works as a meta-surface at the fundamental wavelength, where diffraction is disabled, and as a regular grating at the SH wavelength. By varying either the grating periodicity or the optical wavelength, it is possible to cross the boundary between these two cases. 

The number of diffraction orders in Fig.~1 depends on the angle of incidence $\theta$.
According to the definition, the Wood's anomaly is due to an "uneven distribution of light"\cite{wood1902} between different diffraction orders and manifests itself in the modulation of the intensity of a particular diffraction order when a new diffraction order emerges at $\theta=\theta_W$. The purpose of this study is to observe the nonlinear Wood's anomaly (Fig.~\ref{fig:SHG_diffraction}f) and compare its nonlinear-optical properties with the general case where linear and nonlinear Wood's anomalies coexist (Fig.~\ref{fig:SHG_diffraction}e). 

\section{Experiment}

The sample under investigation is an array of nickel nanodimers deposited on a Si substrate, shown in Fig.~\ref{fig:SHG_diffraction}d. Good heat conductivity of Si ensures the sample damage resilience at high peak optical fluences required for sizeable nonlinear-optical effects. The nickel film was DC sputtered directly on a Si[100] substrate, and then the structure was patterned using electron beam lithography. Nanodisks  with an average diameter of $145$~nm and height of $60$~nm were grouped into dimer cells with the air gap size systematically varying between 15 and 60~nm (15,30,60,15,30,60~nm etc., with an average gap size of $35$~nm). The two symmetry axes of our 2D-array are denoted $p_L$ (longitudinal) and $p_T$ (transverse), as illustrated in Fig.~\ref{fig:SHG_diffraction}~d. Along $p_L$, the longitudinal dimer axis, the distance between the dimers is $\Lambda_L=445$~nm. The triple size of the actual unit cell, caused by the systematically varying size of the gap between the dimers, does not seem to play any role in the present study. Along $p_T$, the periodicity is $\Lambda_T=265$~nm.    

\begin{figure}
    \centering
    \includegraphics[width=1.0\columnwidth]{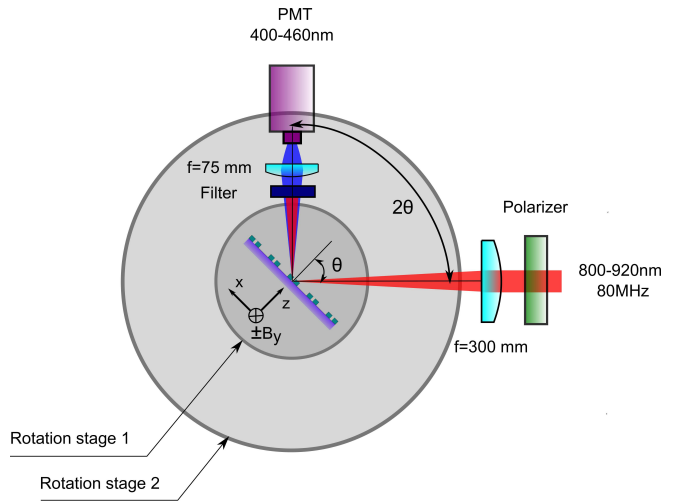}
    \caption{The goniometric setup for angular-dependent magnetic second harmonic measurements on ferromagnetic samples in reflection geometry. The sample and the detection unit (filter, lens and photomultiplier tube) are mounted on two coaxial rotation stages 1 and 2, respectively. The $\sim~100$~mT static magnetic field of a permanent magnet is applied along the $y$-direction.}
    \label{fig:rotStage_Wood}
\end{figure}

The experiment was conducted on a goniometric platform which allows measuring the reflected signals at different angles of incidence, as shown in Fig.~\ref{fig:rotStage_Wood}. 
The fundamental radiation is produced by a mode-locked Ti:Sa laser (Mai-Tai), tunable in the spectral range between 690 and 1040~nm, with $\sim 100$~fs pulse duration and 80~mW average power at 80~MHz repetition rate. The p-polarized (TM) fundamental radiation was focused on the sample surface, yielding a spot of 80~$\mu$m in diameter (FWHM in $y$-direction). The linear reflectivity $R$ was measured using a silicon photodiode. The SH output was spectrally separated by a color BG-39 filter (Schott), collected by a lens and recorded with a photomultiplier tube (Hamamatsu) operating in the photon counting mode. 
In the transverse MOKE configuration (magnetic field $B_y\geq$~100~mT perpendicular to the plane of incidence $xz$, see the reference frame in Fig.~\ref{fig:rotStage_Wood}), the intensity of the reflected SH is modified upon the magnetization reversal in nickel nanodimers giving rise to the magnetization-dependent SH intensity $I_{2\omega}(\pm M)$ in the far field. Originating from the interference of odd and even SH sources, variations in the SH intensity upon reversal of the in-plane magnetization of nickel nanodimers are quantified by the nonlinear magnetic contrast $\rho$: 
\begin{equation} \label{qn:rho}
\rho=\frac{I_{2\omega}(+M)-I_{2\omega}(-M)}{I_{2\omega}(+M)+I_{2\omega}(-M)}\,.
\end{equation}
To study the nonlinear Wood's anomaly we investigated the dependence of SH intensity $I_{2\omega}(\pm M)$ and the associated magnetic contrast on the angle of incidence $\theta$. The linear diffraction anomaly was studied through the angular dependence of reflectivity $R_{\omega}$ at the fundamental wavelength. 
 
Measurements were performed for two orientations of the sample: $E_x\parallel p_T$ and $E_x\parallel p_L$. Fig.~\ref{fig:SHG_diffraction}e-f illustrate the excitation of nonlinear Wood's anomalies for the two orientations. When $E_x\parallel p_L$ ( Fig.~\ref{fig:SHG_diffraction}e), both the linear anomaly $I_\omega^{(-1)}$ and the SH anomaly $I_{2\omega}^{(-2)}$ can be excited at the Wood's angle since $\Lambda_L\geq\lambda/2$, corresponding to the grating regime in Fig.~\ref{fig:SHG_diffraction}b. In contrast, only the SH anomaly $I_{2\omega}^{(-1)}$ is excited at the Wood's angle $\theta_W$ in the case of $E_x\parallel p_T$ (Fig.~\ref{fig:SHG_diffraction}f), which features the transition regime $\lambda/4<\Lambda_T<\lambda/2$ in Fig.~\ref{fig:SHG_diffraction}c. In addition, it has been shown in Ref.~\cite{zubritskaya2015} that the electric field of the plasmon strongly confined in the gap of nickel nanodimers is excited for $E_x\parallel p_L$ in the broad 500-1000~nm spectral range. When $E_x\parallel p_T$, the excitation of this gap plasmon mode is suppressed.       

\section{Results}

\begin{figure}
\centering
\includegraphics[width=0.9\columnwidth]{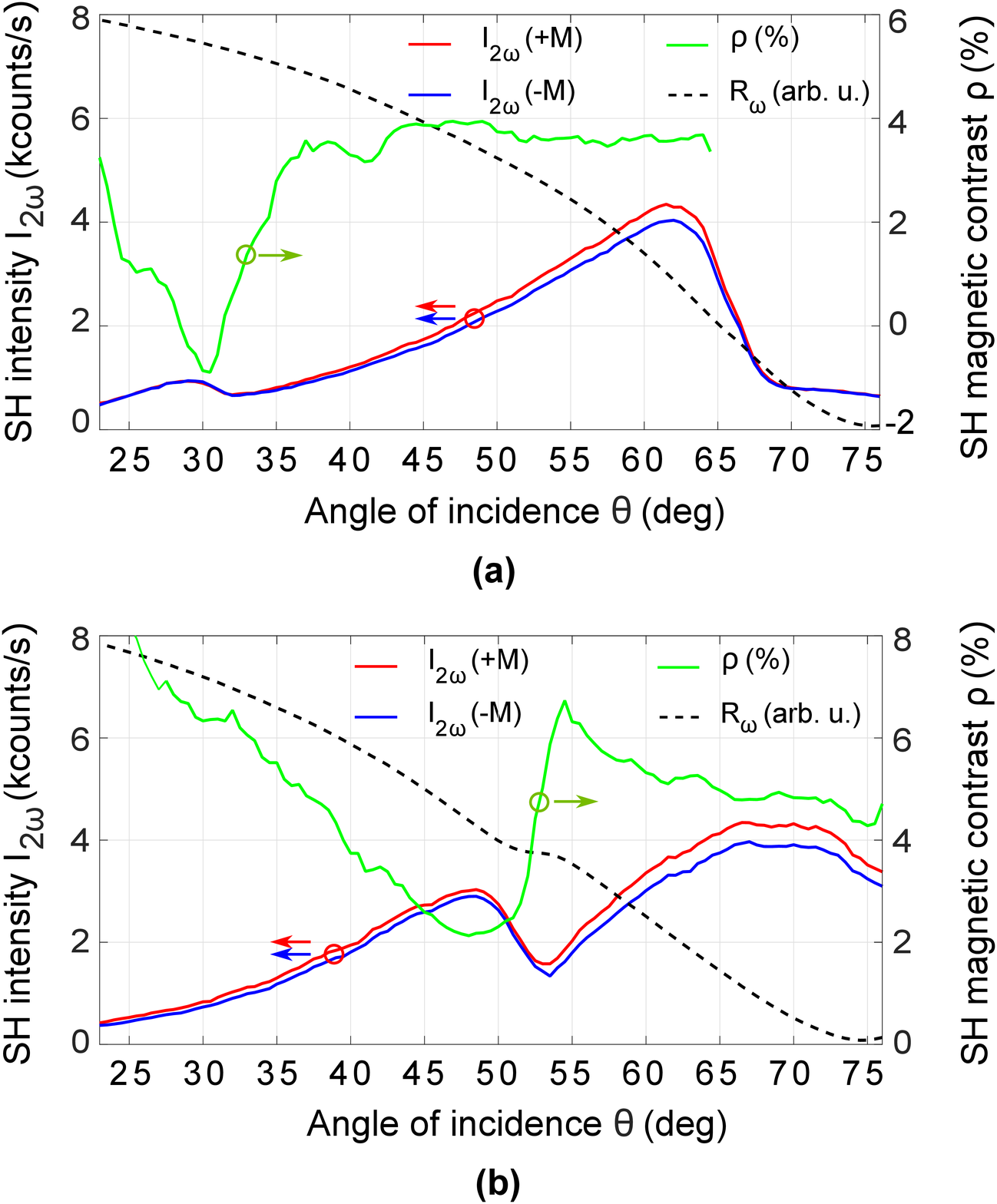}
\caption{The angular dependence of the linear reflectivity $R_\omega$ (dashed black), the SH intensity $I_{2\omega}(\pm M)$ for the two opposite directions of magnetization (solid red and blue), and the SHG magnetic contrast $\rho$ (solid green) at the fundamental wavelength of 820~nm, measured for the case of (a) $E_x\parallel p_T$ and (b) $E_x\parallel p_L$. }
\label{fig:4plot}
\end{figure}

The specular (zero diffraction order $m=0$) linear reflectivity $R_{\omega}$, the magnetization-dependent SH intensity $I_{2\omega}(\pm M)$, and the nonlinear magnetic contrast $\rho$ were measured as a function of angle of incidence $\theta$. Fig.~\ref{fig:4plot}a and Fig.~\ref{fig:4plot}b show the experimental data obtained at the fundamental wavelength $\lambda=$~820~nm for two  orientations of the dimer array, $E_x\parallel p_T$ and $E_x\parallel p_L$, respectively.
 
No significant drop of $R_{\omega}$ has been observed in case of $E_x\parallel p_T$ (Fig.~\ref{fig:4plot}a). 
It means that the linear Wood's anomaly was not observed in this configuration, consistent with the case illustrated in Fig.~\ref{fig:SHG_diffraction}c, where $\Lambda_T<\lambda/2$ and no linear diffraction is allowed except for the zero order. In contrast, in the $E_x\parallel p_L$ geometry with the spatial periodicity $\Lambda_L\approx~445$~nm ($\Lambda_L>\lambda/2$), a small dip in $R_{\omega}$ was observed at $51^{\circ}$ (Fig.~\ref{fig:4plot}b) as a fingerprint of the linear Wood's anomaly excitation around this angle.

In the nonlinear regime, the angular dependence of the SH intensity displays a dip for both orientations: at 32$^{\circ}$ for $E_x\parallel p_T$ and 53$^{\circ}$ for $E_x\parallel p_L$. As compared to the bare silicon substrate, the SH yield from the samples is almost an order of magnitude larger suggesting that the SH is generated predominantly at the nickel-air interface. 
The SHG magnetic contrast $\rho$ also shows the sharp angular features with a magnitude around 5~\% (Fig.~\ref{fig:4plot}) for both orientations of the sample at the same angles. Jumping ahead with conclusions we state that the dips can be interpreted within the framework of the nonlinear Wood's anomaly, to be justified in the following section.


\section{Discussion}

\begin{figure*} 
\centering
 \includegraphics[width=1.0\textwidth]{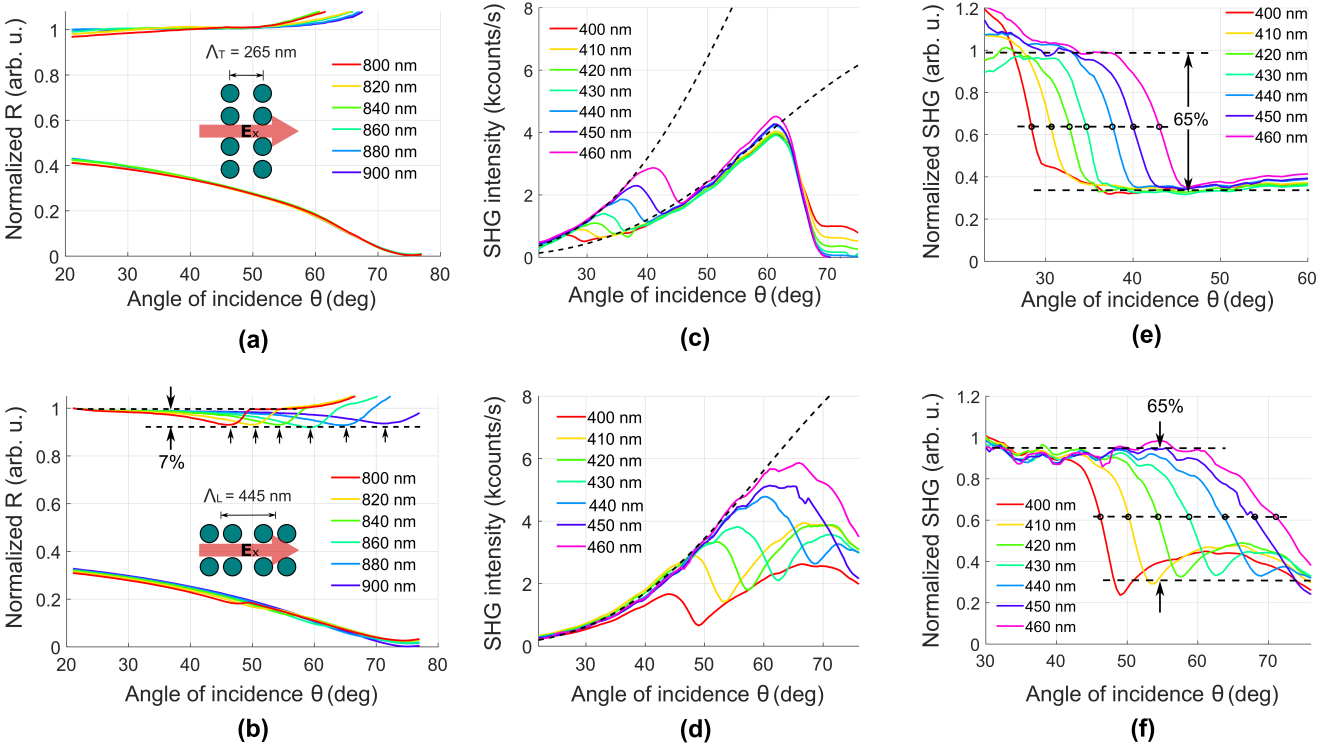}
\caption{The top row (a,~c,~e) shows the analysis of the SHG for $E_x \parallel p_T$. The bottom row (b,~d,~f) presents the data for the case of $E_x \parallel p_L$. (a,~b) Angular dependence of the linear reflectivity at different wavelengths. (c,~d) Angular dependence of SHG intensity at different SH wavelengths. The black dashed lines $\propto\sin^4\theta$ (the two lines in panel (c) possess amplitudes with the ratio of $0.38$) represent an empirical approximation for the angular dependence of SH for a nickel film without diffraction anomalies. (e,~f) The normalized SH signals obtained by dividing SHG intensities in panels (c,~d) by $\propto\sin^4\theta$ demonstrate a 65\% drop of the specular $m=0$ SH intensity upon emergence of new diffraction orders. Small arrows in Fig.~\ref{fig:R}b and crossing points at the threshold 0.6 in Fig.~\ref{fig:R}e-f are used to extract the resonant angles $\theta_W$ for Wood's anomalies.}
\label{fig:R}
\end{figure*}

Further insights in the physical nature and nonlinear properties of the Wood's anomalies can be obtained from spectral measurements. We performed angular measurements of the linear reflectivity and magnetization-dependent SH output at various fundamental wavelengths. As a result, we observed a pronounced wavelength-dependence of angular features in the linear reflectivity, SH intensity and magnetic contrast. The data for two sample orientations are summarized in Fig.~\ref{fig:R}.

In Fig.~\ref{fig:R}a-b we inspect the specular linear reflectivity $R_{\omega}$ at different wavelengths.
In these plots, the reflectivity curves at the bottom ($R\leq$~0.4) are the raw data, while those at the top ($R>$~0.8) are normalized to the measured reflectivity of the bare silicon substrate. 
The high values of the normalized $R_{\omega}$ at angles beyond $60^{\circ}-70^{\circ}$ are artifacts originating from the normalization procedure. 

In the $E_x\parallel p_T$ configuration (Fig.~\ref{fig:R}a), the reduced spatial periodicity $\Lambda_T=265$~nm prohibits diffraction.
In the $E_x\parallel p_L$ configuration (Fig.~\ref{fig:R}b), within the wavelength range of $800-920$~nm and angles of incidence $25^{\circ}-75^{\circ}$, diffraction is possible and the Wood's anomaly manifests itself in reflectivity minima shifting towards larger $\theta$ upon the increase of the wavelength. The drop of reflectivity is about $7\%$ for all wavelengths.  

A considerably richer picture can be recovered from the SH angular spectra (Fig.~\ref{fig:R}c-d). The dependence of SH output on the angle of incidence and wavelength in the $E_x\parallel p_T$ geometry exhibits a strong SHG peak at $\theta\approx 61.5^{\circ}$, independent on the wavelength. In contrast to that, the small-angle SH peak 
shifts towards larger angles of incidence for longer fundamental wavelengths. A similar dispersive shift of the SH features is observed in the other geometry ($E_x\parallel p_L$), albeit at different angles. As in the case of linear Wood's anomaly we attribute this SHG minimum to the appearance of the new diffraction order. Apart from the resonant angle for the nonlinear Wood's anomaly the SHG intensity follows the   
empirical background $\propto\sin^4\theta$, see Fig.~\ref{fig:R}c-d.  
The normalization to this background in Figs.~\ref{fig:R}e-f clearly demonstrates a pronounced decrease of the specular SHG intensity at large angles of incidence for a multitude of fundamental wavelengths. 
The $65\%$ drop of the SHG intensity is almost the same for both the purely nonlinear regime (Fig.~\ref{fig:R}e) and the mixed linear-nonlinear Wood's anomaly scenario (Fig.~\ref{fig:R}f). Furthermore, the percentage of the energy pumped into the second-order $m_{nl}=-2$ in the configuration $E_x\parallel p_L$ is not consistent with the mere $7\%$ drop of the reflectivity $R_{\omega}$ measured for the linear diffraction anomaly (see Fig.~\ref{fig:R}b). Based on these observations, we speculate that in the mixed case the nonlinear Wood's anomaly dominates. Moreover, in this geometry, the LSPRs can be excited, characterized by an enhancement of the electric field in the dimer gap. However, the absolute SHG yield in both parallel and longitudinal experimental configurations is almost identical suggesting that possible contributions of LSPRs to SHG can be neglected.  
 
As such, for both experimental configurations, our results evidence the ``uneven distribution of light" between SH diffraction orders, extending the Wood's picture into the nonlinear-optical domain and demonstrating the immense sensitivity of SH spectroscopy to these effects.

\begin{figure}
\centering
\includegraphics[width=1\columnwidth]{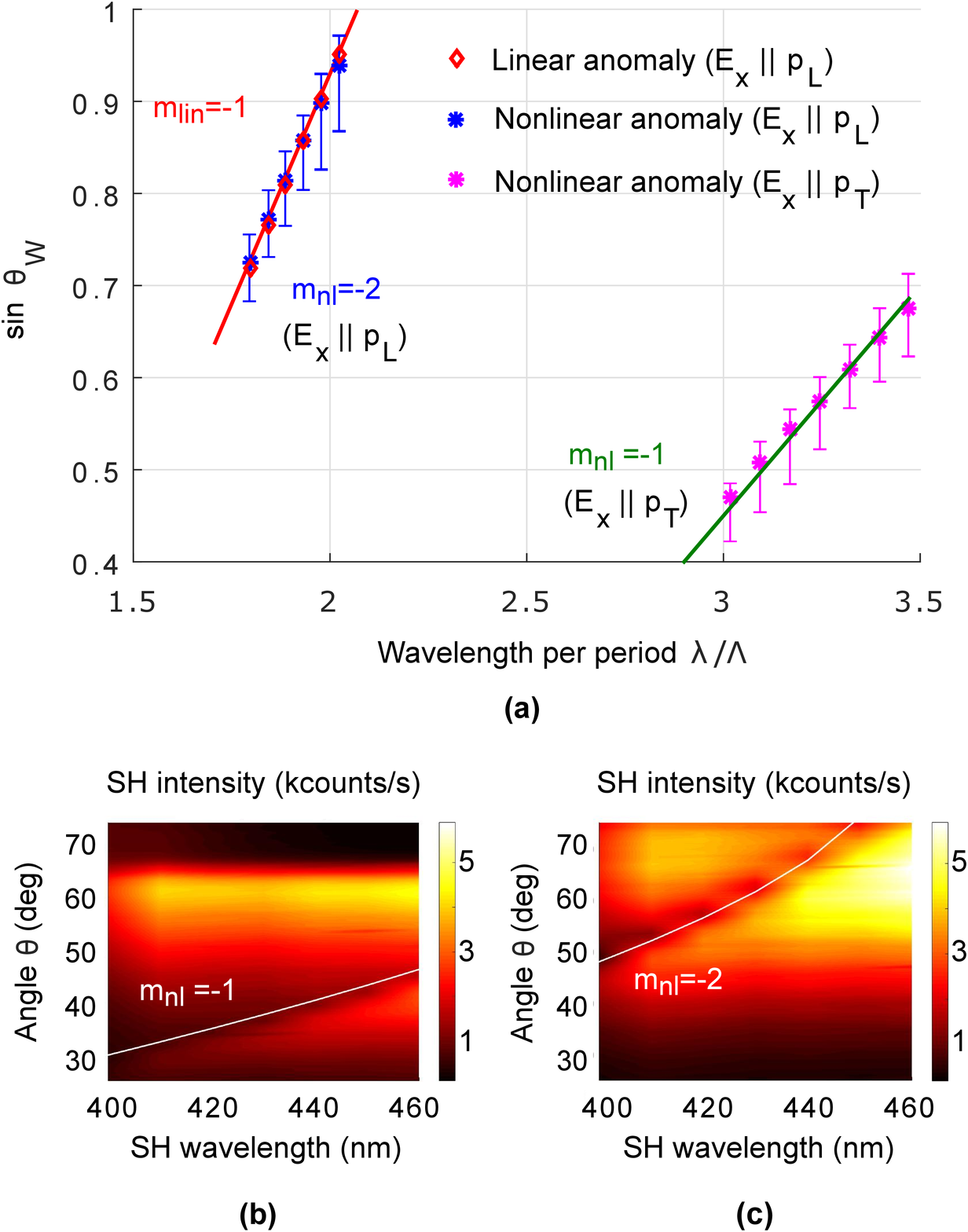}
\caption{(a) The lines show the theoretical dispersion of Wood's anomalies excited in the investigated array. The points show the $\sin\theta_W$ extracted experimentally, where the angles $\theta_W$ are indicated by arrows in Fig.~\ref{fig:R}b and empty dots in Fig.~\ref{fig:R}e-f. (b) For $E_x \parallel p_T$, the first-ordered nonlinear diffraction anomaly $m_{nl}=-1$ (white line) can be excited, matched with the dip (dark region) in the angular dependence of SHG. (c) For the case when $E_x \parallel p_L$, the SHG minima (dark region) fit in the nonlinear Wood's anomaly with $m_{nl}=-2$ (white line).}
\label{fig:dispersion}
\end{figure} 

The spectral dependence of resonant angles $\theta_W$ is extracted from Fig.~4, with values marked by arrows in Fig.~\ref{fig:R}b and empty dots in Fig.~\ref{fig:R}e,f. These resonant angles can be grouped into three different sets: Fig.~\ref{fig:dispersion}a shows that $\sin\theta_W$ scales linearly with the normalized fundamental wavelength $\lambda/\Lambda$. In agreement with an intuitive picture of Fig.~\ref{fig:SHG_diffraction}e, linear and nonlinear diffraction anomalies for $E_x\parallel p_L$ configuration occur at the same angle. We did not introduce error bars for the linear Wood's anomaly as the only reasonable way to extract its resonance angle $\theta_W$ was from the asymmetric minima of the normalized linear reflectivity in Fig.~\ref{fig:R}b. For the nonlinear case the error bars account for the finite angular width of the threshold-like drop of the normalized SH intensity from 1 to 0.35. 

Quantitative analysis of linear fits and experimental error bars appears to be quite instructive. The condition for the linear Wood's anomaly, i.e. the diffraction beam of order $m_{lin}=-1$ at the fundamental frequency propagating along the surface, directly follows from Eq.~(\ref{linear_PM}): 
\begin{equation} \label{qn:k-match}
n_{\rm eff}(\omega)k_0=-k_0\sin\theta_W - m_{lin}G\,.
\end{equation} 
Here $n_{\rm eff}(\omega)$ is an effective index of the diffraction beam accounting for the fact that light propagating along the surface also interacts with nickel nanodisks. From the data presented in Fig.~\ref{fig:dispersion}a, one obtains $m_{lin}=-1$ and $n_{\rm eff}(\omega)=1.07$.  
The excitation condition for the nonlinear Wood's anomaly for the diffraction order $m_{nl}$ follows from Eq.~(\ref{nonlinear_PM}): 
\begin{equation} \label{qn:k-match_nl}
n_{\rm eff}(2\omega)2k_0=-2k_0\sin\theta_W - m_{nl}G\,.
\end{equation}
Linear fitting for $E_x\parallel p_L$ configuration provide $m_{nl}=-2$ and $n_{\rm eff}(2\omega)=1.07$. Taking into account the error bars we obtain $n_{\rm eff}(2\omega)=1.07^{+0.07}_{-0.03}$. For the $E_x\parallel p_T$ configuration the fit parameters are $m_{nl}=-1$ and $n_{\rm eff}(2\omega)$=1.05 (or $n_{\rm eff}(2\omega)=1.05^{+0.06}_{-0.03}$ from the analysis of error bars). 

Our experimental observation of $n_{\rm eff}>1$ is consistent with the results by Lezec and Thio~\cite{lezec2004}, who found $n_{\rm eff}\simeq 1.1$ for a variety of plasmonic and non-plasmonic periodic structures. 
For our sample, the nickel content (the filling fraction of the surface) is 29\%, sufficiently low so that conventional thin-film SPPs cannot be excited. In contrast, signatures of propagating dipole-dipole coupled modes have been observed on periodic arrays of gold nanodisks~\cite{CrozierOE2007}. Excitation of LSPRs in the gap between the nanodisks may also contribute to $n_{\rm eff}(\omega)>1$. However, values of $n_{\rm eff}=1.07$ for $E_x\parallel p_L$ and $n_{\rm eff}=1.05$ for $E_x\parallel p_T$ configurations coincide within the error bars. As such, it is impossible to speculate about possible contributions of LSPRs and propagating dipole-dipole coupled modes to $n_{\rm eff}>1$. It is likely that the spatial localization of the  composite diffracted evanescent wave (CDEW), excited at and propagating along the surface of periodic nanostructures under conditions of Wood's anomaly, would provide the correct interpretation~\cite{lezec2004}.  
CDEW propagates in the air and possesses the non-zero imaginary $k_z$ responsible for an exponential decay in the $z$-direction and $k_x=n_{\rm eff}k_0=\sqrt{k^2_0+k^2_z}>k_0$. Assuming $n_{\rm eff}=1.05$ and $\lambda$=800~nm we obtain a reasonable estimation of 400~nm for the localization length of CDEW electric field in the $z$-direction.

\begin{figure}
\centering
\includegraphics[width=1.0\columnwidth]{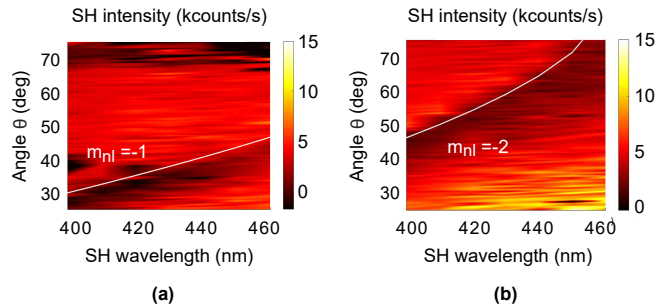}
\caption{The false-color images of the SHG magnetic contrast for (a) $E_x\parallel p_T$, where the white line shows the theoretical dispersion for the second-harmonic anomaly $m_{nl}=-1$, and (b) $E_x\parallel p_L$, where the white line marks the dispersion of $m_{nl}=-2$. }
\label{fig:rho}
\end{figure}

Fig.~\ref{fig:dispersion}b-c represent the SHG data as false-color two-dimensional maps together with the theoretical dispersion by Eq.~(\ref{qn:k-match_nl}). In both configurations, the minimum of the specular SH output is well correlated with the calculated dispersion of nonlinear Wood's anomalies. The magneto-optical fingerprint of the nonlinear Wood's anomaly is exemplified in similar false-color maps of SHG magnetic contrast in Fig.~\ref{fig:rho}: the drop of the magnetic contrast $\rho$ (dark area, cf.~Fig.~\ref{fig:4plot}) again correlates with the calculated dispersion of the nonlinear diffraction anomalies. We conclude that nonlinear magneto-optics offers an additional degree of freedom for experimental characterization of the nonlinear Wood's anomaly on periodically nanostructured ferromagnetic samples. Interestingly, this result is inherently consistent with our previous findings reporting on the dependence of the nonlinear SPP excitation in thin metal films on the angle of incidence in Kretschmann configuration and an external magnetic field ~\cite{razdolski2016,Temnov2016}. From the nonlinear-optical point of view, the similarity between these seemingly different phenomena could be related to strong variations of an electric field at the double frequency $2\omega$ induced by (i) interference of its odd and even components and/or (ii) the nonlinear phase-matching conditions. The physical origin of this similarity represents an interesting question in nonlinear photonics and requires further investigation.

\section{Conclusions}

In this work, we have investigated the impact of Wood's anomalies on the SH output and nonlinear MOKE on a two-dimensional arrays of nickel nanodimers. We measured the angular spectra of the linear reflectivity and the magnetic SH intensity $I_{2\omega}(\pm M)$ for a series of fundamental wavelengths and two orientations of the sample. A detailed comparison of these two data sets highlighted the relevant linear and nonlinear excitation mechanisms of Wood's anomaly.

Our results suggest that the linear Wood's anomaly has little influence on the SH yield. The decrease of SHG intensity $I^{(0)}_{2\omega}$ upon the emergence of the new $m_{nl}=-1$ or $m_{nl}=-2$ diffraction orders in the nonlinear regime is $65\%$, i.e. an order of magnitude larger than $7\%$ reflectivity variations in its linear counterpart. The SHG magnetic contrast also shows a fingerprint of the nonlinear diffraction anomaly, which is the main factor contributing to the $\sim 5\%$ angular variation of the SHG magnetic contrast $\rho$ around the Wood's anomaly. Nonlinear magneto-optical effects can thus be utilized as a measure of the efficiency of diffraction gratings in periodic arrays of magnetic nanoparticles. In the investigated array of nickel nanodimers the large difference in the lattice period between transverse and longitudinal configurations determines the character of Wood's anomalies. In future, possible contributions of LSPRs in nanodimers resulting in the excitation of SLRs upon interference with Wood's anomalies might be identified in similar experiments using structures with identical longitudinal and transverse lattice periods and a series of different gap sizes. Furthermore, an interplay between the nonlinear magneto-optical Wood's anomaly and the Peierls transition (when a series of nanodimers collapses in an equidistant chain of ferromagnetic nanodisks) seems to be nontrivial. Wood's anomaly is known to be enhanced on metallic gratings supporting propagating SPP modes. It would be interesting to perform similar measurements on periodic arrays of sub-wavelength holes aiming at a complimentary magneto-optical view on the phenomenon of extraordinary optical transmission~\cite{Ebbesen1998,lezec2004} in the nonlinear-optical regime.

\begin{acknowledgments}
The dimer patterning was performed at the Center for Functional Nanomaterials, Brookhaven National Laboratory, supported by the U.S. Department of Energy, Office of Basic Energy Sciences, under Contract No. DE-SC0012704. The authors thank Rimantas Brucas, Henry Stopfel and Tobias Warnatz for their help involving the preparation and patterning of the sample, and Martin Wolf for his support.\\
The authors acknowledge Deutsche Forschungsgemeinschaft (AL2143/2-1), Agence Nationale de la Recherche for financial support under grant ''PPMI-NANO'' (ANR-15-CE24-0032 and DFG SE2443/2), Strat\'egie internationale ''NNN-Telecom'' de la R\'egion Pays de La Loire, the Knut and Alice Wallenberg Foundation project 'Harnessing light and spins through plasmons at the nanoscale' (2015.0060), the Swedish Research Council and the Swedish Foundation for International Cooperation in Research and Higher Education. This work is part of a project which has received funding from the European Union's Horizon 2020 research and innovation programme under grant agreement no.~737093.
\end{acknowledgments}

\bibliographystyle{unsrt}

\end{document}